\documentclass[pdflatex,iicol]{sn-jnl}
\usepackage[square,numbers]{natbib}
\usepackage[normalem]{ulem}

\jyear{2021}%

\theoremstyle{thmstyleone}%
\theoremstyle{thmstyletwo}%

\theoremstyle{thmstylethree}%

\iftrue
\usepackage{etoolbox}
    \makeatletter
    \patchcmd{\ps@headings}
    {Springer Nature 2021 \LaTeX\ template}
    {}
    {}
    {}
    \patchcmd{\ps@headings}
    {Springer Nature 2021 \LaTeX\ template}
    {}
    {}
    {}
    \patchcmd{\ps@titlepage}
    {Springer Nature 2021 \LaTeX\ template}
    {}
    {}
    {}
    \makeatother
\fi

\begin{document}
\title[Collaborative Cluster Configuration for Distributed Data-Parallel Processing]{Collaborative Cluster Configuration for Distributed Data-Parallel Processing: A Research Overview}

\author[1,2]{\fnm{Lauritz} \sur{Thamsen}}\email{\{firstname.lastname\}@tu-berlin.de}

\author[1]{\fnm{Dominik} \sur{Scheinert}}

\author[1]{\fnm{Jonathan} \sur{Will}}

\author[1]{\fnm{Jonathan} \sur{Bader}}

\author[1]{\fnm{Odej} \sur{Kao}}

\affil[1]{\orgname{Technische Universität Berlin}, \orgaddress{\city{Berlin}, \country{Germany}}}

\affil[2]{\orgname{University of Glasgow}, \orgaddress{\city{Glasgow}, \country{United Kingdom}}}


\abstract{
Many organizations routinely analyze large datasets using systems for distributed data-parallel processing and clusters of commodity resources.
Yet, users need to configure adequate resources for their data processing jobs.
This requires significant insights into expected job runtimes and scaling behavior, resource characteristics, input data distributions, and other factors.
Unable to estimate performance accurately, users frequently overprovision resources for their jobs, leading to low resource utilization and high costs.\\
In this paper, we present major building blocks towards a collaborative approach for optimization of data processing cluster configurations based on runtime data and performance models.
We believe that runtime data can be shared and used for performance models across different execution contexts, significantly reducing the reliance on the recurrence of individual processing jobs or, else, dedicated job profiling.
For this, we describe how the similarity of processing jobs and cluster infrastructures can be employed to combine suitable data points from local and global job executions into accurate performance models.
Furthermore, we outline approaches to performance prediction via more context-aware and reusable models.
Finally, we lay out how metrics from previous executions can be combined with runtime monitoring to effectively re-configure models and clusters dynamically.
}




\maketitle

\section{Introduction}
\label{sec:introduction}

Numerous organizations work with large volumes of data, be it to recommend content to millions of users~\cite{das2007google}, to identify disorders by comparing terabytes of genomic data, to monitor environmental conditions using large-scale distributed sensor networks~\cite{aberer2007infrastructure}, or to detect fraudulent behavior in millions of business transactions~\cite{chan1999distributed}.
For this, businesses, sciences, and municipalities often deploy data-intensive applications that run on distributed data-parallel processing systems and large-scale virtualized computing infrastructures.

Distributed data-parallel processing systems provide high-level programming abstractions, efficient data-parallel operator implementations, and distributed task communication for developing massively parallel processing jobs for compute clusters.
Prominent example systems include MapReduce~\cite{dean2008mapreduce}, Spark~\cite{zaharia2010spark}, and Flink~\cite{carbone2015apache}.

Virtualization abstracts the technical and configuration details of the computer hardware, allowing users to flexibly provision virtual resources for their data processing jobs without detailed knowledge of the underlying infrastructure.
However, the challenge of capacity planning for a specific processing job remains:
Which type of resource should be selected for a job?
How many of such processing units should be allocated?
And which system-specific resource configurations (task parallelism, memory allocations, etc.) should be set for each unit?
Determining adequate resource configurations requires significant knowledge and is often difficult even for expert users~\cite{rajan2016perforator,witt2019predictive,bader2021tarema}.
Efficient cluster configurations need to be chosen out of those that fulfill possibly multiple objectives and constraints as defined by users or service level agreements (SLAs).
These can, for instance, define expectations for the runtime of jobs as well as monetary and environmental costs of executions.
A good decision therefore commonly calls for a thorough understanding of a job’s runtime behavior to estimate its scalability, performance, and costs on a particular type of resource prior to executing the job.
However, this behavior depends on a number of factors, and users often have only limited insights into these before jobs are executed.
Moreover, despite all its benefits, virtualization can introduce unexpected overheads and fluctuations to a job's performance at runtime.

At the same time, infrastructure providers develop and offer ever-increasing numbers of available resource configurations, so users must decide among dozens to hundreds of virtual machine types and specify the desired scale-out in public clouds.
Amazon EMR\footnote{\url{https://aws.amazon.com/emr/}}, for example, a cloud platform for big data processing, currently offers 128 different virtual machine types, from general-purpose machines to decidedly specialized resources.
For dedicated clusters, on the other hand, the situation is often similar.
Scientists typically have access to multiple clusters, each providing access to different resources.
The authors of this article, for instance, can use more than half a dozen clusters, of which multiple have diverse nodes and processing units.

Given the difficulty of selecting resources and the large number of options, users tend to over-provision resources for their jobs to ensure that performance expectations are met. For example, studies of production clusters at large companies regularly present an aggregated resource utilization of well under 50\%~\cite{liu2011measurement,delimitrou2014quasar,cheng2018characterizing}, while resource reservations are often several times higher. Take, for instance, the analysis of a data analytics cluster at Twitter~\cite{delimitrou2014quasar}, which presents an aggregated CPU utilization of barely 20\% while close to 80\% of the CPU resources were reserved.
This wasteful usage of reserved resources leads to poor overall cluster utilization as well as long waiting times for users, who are then not able to access the unused but blocked resources.

Many studies addressed this problem by developing and proposing performance models for more automated and effective resource management~\cite{venkataraman2016ernest,sidhanta2016optex,thamsen2016bell,verbitskiy2018cobell,witt2019predictive,alipourfard2017cherrypick,hsu2018micky,hsu2018arrow,chao2018gray,shah2019quick,al2020gray}.
However, the proposed approaches either rely on the availability of historical runtime data or, else, on dedicated profiling runs, which are frequently unavailable or, else, prohibitively costly in real-world settings.

The approach described in this paper addresses scenarios in which historical data is not sufficiently available locally, and the inherent overhead of dedicated profiling runs is also not an option. The main idea is to follow a collaborative approach to cluster configuration for distributed data-parallel processing, envisioning that users share historical job runtimes and performance models with each other to use this information across different execution contexts.
This enables performance estimations and, in turn, model-based cluster configuration optimization, even in situations in which users have not executed a specific job repeatedly already.
Therefore, cluster hours, operational costs, and consumed energy can potentially be reduced for jobs.

In order to develop suitable methods for this collaborative vision, we focused our efforts on answering the following three research questions:

\begin{enumerate}
    \item How can runtime data be shared between users and data points from similar jobs be used for performance models in different execution contexts?
    \item How can performance models or their components be designed to be more context-aware and reusable to support their application across different execution contexts?
    \item How can performance models and cluster configurations be adjusted efficiently at runtime as more data on a job's actual performance becomes available?
\end{enumerate}

Over the last years, we have developed and tested several major building blocks towards our collaborative approach to cluster configuration in our projects in the research center BIFOLD\footnote{\url{https://bifold.berlin/}}, proposing answers to our research questions.
In this paper, we present a holistic picture by connecting the existing methods, highlight and discuss our main findings, and briefly outline our next steps.
We first provide an overview of state of the art (Section~\ref{sec:related_work}).
We then present our collaborative approach to cluster configuration for distributed data-parallel processing (Section~\ref{sec:approach}).
Afterwards, the paper summarizes our previous results and future plans for realizing our vision (Section~\ref{sec:results}).
Finally, the paper concludes with a short summary (Section~\ref{sec:summary}).

\section{Related Work}
\label{sec:related_work}
This section discusses related works on performance modeling for efficient distributed data-parallel processing.

\paragraph{White- \& Grey-Box Performance Modeling:}
We first highlight approaches that focus on a specific data processing framework or a class of algorithms and are thus completely (white-box) or partially (gray-box) framework-dependent.

To start with, Apache Spark's multi-stage execution structure is explicitly utilized in~\cite{wang2015performance}.
Using runtime information obtained from sample runs, the behavior of individual stages is learned and used for performance prediction.\\
A similar method is OptEx~\cite{sidhanta2016optex}, which incorporates information about the cluster size, the number of iterations, the input dataset size, and certain model parameters into an analytical model.\\
Doppio~\cite{ZhouRFSRC18doppio} analyzes the relation between I/O access and computation to build its Spark-specific prediction model, which can be applied on shuffle-heavy and iterative Spark jobs.\\
With regards to gray-box approaches, one method is proposed in~\cite{chao2018gray}, where the authors train specialized regression models for each stage of a Spark application.
They further take into account cluster hardware information and resource allocation mechanisms of Spark.\\
PERIDOT~\cite{shah2019quick} derives an analytical model that combines knowledge of Spark's data partitioning mechanisms with information obtained from a short profiling phase.
The thereby obtained model allows for accurate performance predictions.\\
Another gray-box method is proposed in~\cite{al2020gray},
where one model is used to predict the input sizes of stages, considering several Spark application parameters, whereas another model utilizes those predictions to estimate the runtime of tasks.

The presented works make detailed assumptions about the inner functioning of specific frameworks, which makes them completely or partially framework-dependent.
In contrast, our work is applicable to not just a single data processing framework, but the general class of distributed data-parallel processing.

\paragraph{Black-Box Performance Modeling:}

Black-box approaches for runtime performance estimation are framework-independent and can, in principle, be used with various systems.

Some approaches aim to iteratively find a good cluster configuration by considering runtime information from prior iterations.
Once another search iteration yields no significant improvement, the approaches settle on a near-optimal state~\cite{alipourfard2017cherrypick,hsu2018micky,bilal2020vanir}.\\
CherryPick~\cite{alipourfard2017cherrypick} selects near-optimal configurations in the cloud with low overhead and high accuracy.
The authors achieve this by using Bayesian optimization to direct the profiling process until a good enough solution is found.\\
Micky~\cite{hsu2018micky} uses a collective optimizer that simultaneously profiles several workloads to improve modeling efficiency.
The authors reformulate the original problem as a multi-armed bandit problem to balance exploration and exploitation.\\
Another example is Vanir~\cite{bilal2020vanir}, which first finds an initial configuration for profiling runs by using a heuristic method, and then utilizes a performance model to improve the configuration in the production runs progressively.
In both steps, transfer learning is used if possible.\\
The disadvantage of these iterative approaches is that they rely on dedicated profiling in order to gather sufficient data.
This is associated with overhead in both time and cost.

Aside from iterative approaches, there are also methods that make use of existing runtime data~\cite{venkataraman2016ernest,rajan2016perforator}.
In case runtime data from historical runtime data is not yet available, it can initially be generated through dedicated profiling runs, which are usually selected to provide an overview of the cluster configuration search space.\\
Ernest~\cite{venkataraman2016ernest} employs a parametric model for estimating the performance of machine learning applications based on dataset sizes and the number of machine instances, yet is strictly limited to individual instance families.\\
In contrast, PerfOrator~\cite{rajan2016perforator} uses analytical models for the underlying framework, calibration queries, and non-linear regression on profile runs to tackle the aforementioned limited prediction capabilities.
Thus, the system can predict job runtimes, costs, as well as resource usage.
\\
Lastly, Leis et al.~\cite{leis2021towards} propose a modeling approach that estimates the runtime and cost of analytical database queries for specific machine instances in a cloud environment.

Our work extends the aforementioned methods by explicitly enabling the sharing of runtime data and performance prediction models among many collaborators.
Moreover, we put emphasis on the transferability and adaptivity of models across different infrastructures and target processing jobs.
With that, we reduce the reliance on profiling runs and recurring job executions.

\section{General Idea}
\label{sec:approach}

\begin{figure}[b!]
    \includegraphics[height=0.85\textheight, width=\columnwidth, keepaspectratio]{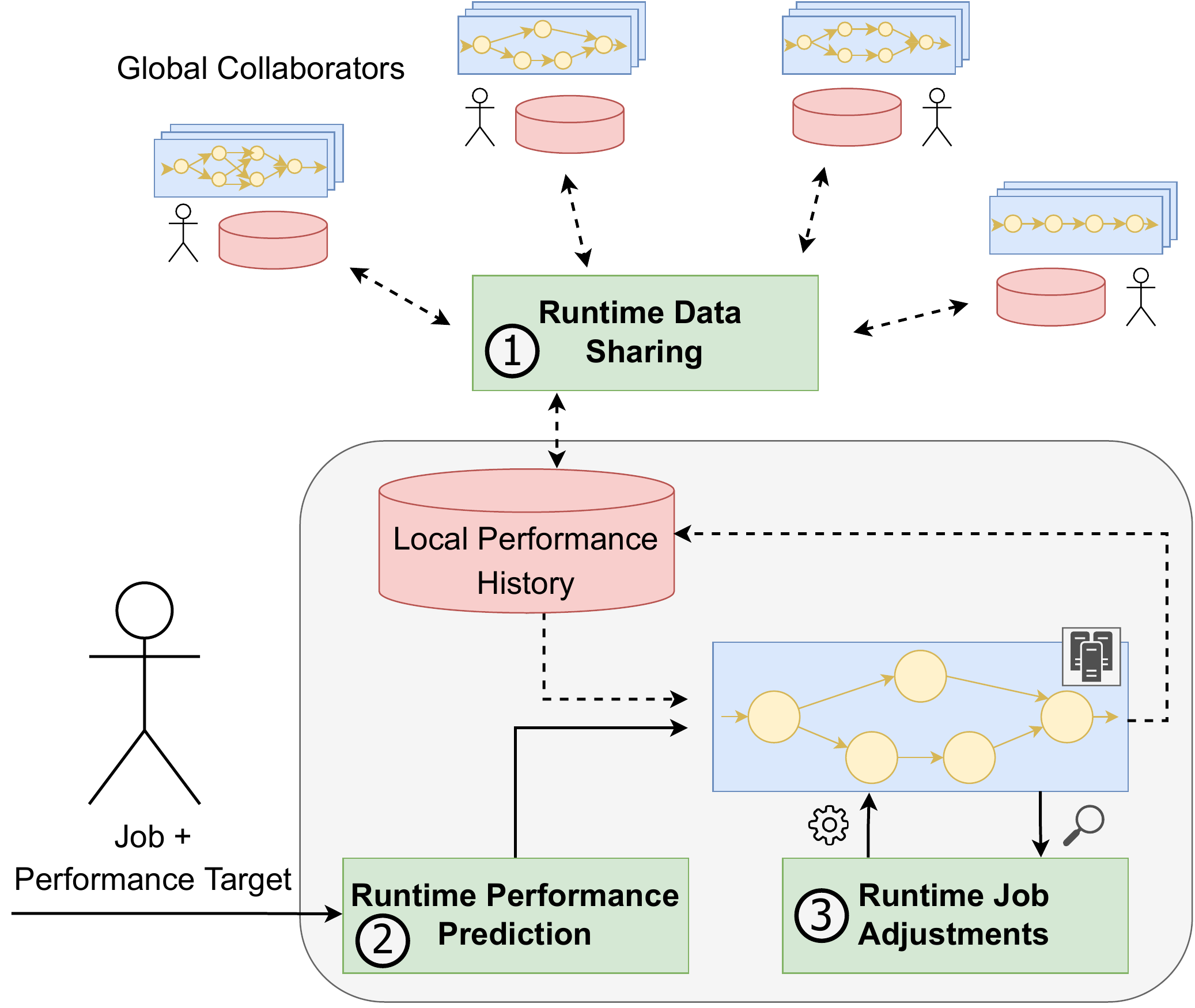}    \caption{Overview of our collaborative process, which benefits from \raisebox{.5pt}{\textcircled{\raisebox{-.9pt} {1}}} runtime data sharing, an important enabler for \raisebox{.5pt}{\textcircled{\raisebox{-.9pt} {2}}} runtime performance prediction and \raisebox{.5pt}{\textcircled{\raisebox{-.9pt} {3}}} continuous runtime job adjustments.}
    \label{fig:approach_overview}
\end{figure}

An overview of our collaborative approach to configuring cluster resources for distributed data-parallel processing is provided by~\autoref{fig:approach_overview}.
Central aspects are \emph{runtime data sharing}, \emph{runtime performance prediction}, and \emph{runtime job adjustments}.
Collaborators are enabled to \raisebox{.5pt}{\textcircled{\raisebox{-.9pt} {1}}} share runtime data alongside code in decentralized repositories, \raisebox{.5pt}{\textcircled{\raisebox{-.9pt} {2}}} train context-aware models with the globally gathered runtime information and tune them for the current local context, and \raisebox{.5pt}{\textcircled{\raisebox{-.9pt} {3}}} continuously update performance models and adjust resource configurations during the execution of a job.
These aspects are further discussed in the following.

\subsection{Runtime Data Sharing}

Our aim is to enable sharing of runtime data among different users to train performance models collaboratively.
However, the shared runtime data can stem from significantly different execution contexts.
We approach this problem by taking advantage of the similarity of different cluster environments and jobs, which can be captured in infrastructure and job performance profiles.
More specifically, we can use information readily available on nodes and network interfaces, such as performance metrics presented on websites of manufacturers and published benchmarks, as well as command-line accessible resource information.
Additionally, repositories can store profiles of infrastructure components and processing jobs, assuming compiled data-parallel execution plans which can be used for finding matches, to be filled over time with measurements of selected microbenchmarks.
To further abstract individual resources, clustering methods can automatically group resources with similar static and dynamic characteristics~\cite{bader2021tarema}.

Additional challenges are introduced when jobs do not use resources in isolation, but share access and potentially interfere with each other, impeding individual job performance often significantly~\cite{verbitskiy2018cobell,thamsen2021mary}.

Job profiles can be used in conjunction with infrastructure profiles, allowing to make sensible selections among different resources, determining specific scale-outs of jobs, and taking co-location of multiple different jobs into account.

\subsection{Runtime Performance Prediction}

Utilizing shared runtime data, we aim to develop performance prediction models that are aware of execution contexts and hence reusable.
We argue that this requires the identification of suitable modeling approaches in the first place, flexible model selection strategies based on data availability, and efficient cross-context transfer of already trained models for local fine-tuning.

In general, the runtime performance models need to be fairly agnostic to the specific tools used for distributed data-parallel processing in order to be widely applicable, yet can still assume certain patterns of distributed data-parallel processing.

For any type of job, there exist most likely different valid performance models, of which the best performing is expected to be found depending on the available training data and prediction task.
Exploiting such problem-specific variability demands little overhead from an initial model selection procedure, which can be realized collaboratively as well if selection strategies or even trained models are shared among users.
For scenarios in which not just runtime data but also prediction models are shared, we need the models to effectively capture the execution context they were produced in, including information on the resources and system configurations used.

The models should prioritize similar and local runtime data where it is available as it can be expected to be more indicative of future executions.
Approaches for this prioritization are, on the one hand, pre-trained models which can be tuned on local data and, on the other hand, explicit representations of different contexts for comparability and improved data selection.

\subsection{Runtime Job Adjustments}

Even robust prediction models can produce inaccurate estimates at runtime, because there is inherent variance due to resource sharing and potential failures, as well as changed input data.
It is therefore necessary to continuously monitor the execution of a job, automatically update performance models as a result of newly available data, and as a consequence adjust both resource and system configurations at feasible points in time.

We see significant potential in using data collected as a job runs to compare it to previously recorded executions dynamically.
This dynamic runtime data can include runtimes and resource utilization of specific parts of jobs, data placement information, and data access patterns.
Such data can further be given weight dynamically.
The dynamic runtime data can then be joined with static job configuration information, allowing to compare job executions comprehensively.

With regards to performance model updates, we can distinguish between long-term degradation of performance estimations and extraordinary events.
Depending on this, but also the model used, different approaches can be used to implement efficient model updates.
Complex models can be partially updated immediately and scheduled for re-training from scratch later.
On the other hand, small and specialized models often allow for quick re-training.

In case of planned re-configurations, systems should take the overheads for configuration adjustments into account as updates at runtime are often quite costly.
Such adjustments should be implemented at points where the migration overhead is comparatively low, yet where the re-configuration is still expected to have a substantial impact.
Historical runtime data, specialized performance models, and time series forecasting techniques seem well-suited for identifying promising points for dynamic adjustments.

\section{Main Results}
\label{sec:results}
In this section, we summarize our main results so far and provide a brief outlook toward future work.

\subsection{Runtime Data Sharing Through Open and Decentralized Repositories}
\label{sec:results_data_sharing}

First, we present strategies that enable users to benefit from runtime data that was generated outside of their own execution environment.

\paragraph{Results Overview}

\begin{figure}[htb!]
    \includegraphics[height=0.85\textheight, width=\columnwidth, keepaspectratio]{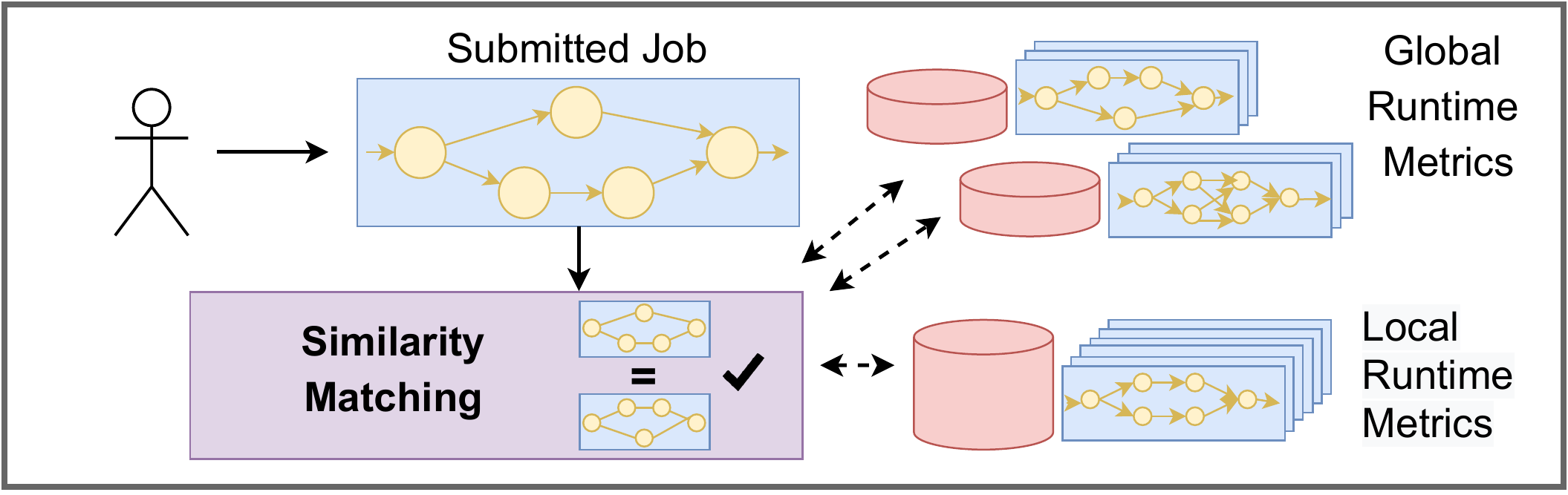}
    \caption{Matching runtime metrics of local and global historical job executions to submitted data processing jobs. Combining runtime metrics of similar local and global historical job executions to use for performance models.}
    \label{fig:sharing_overview}
\end{figure}

In several of our prior works~\cite{koch2017smipe,will2020towards,will2021c3o,will2021training,scheinert2021potential}, we discussed the idea of exploiting similarities between different jobs and their executions, cultivating runtime data in a collaborative manner among numerous users and thereby improving the prediction capabilities of individual users.
This includes decentralized system architectures for sharing context-aware runtime metrics, as well as similarity matching between jobs.
An abstract depiction of this idea can be seen in~\autoref{fig:sharing_overview}.

\paragraph{Central Findings}

Worth highlighting is C3O~\cite{will2021c3o}, where we took first steps toward system architectures that organize context-aware runtime data and performance model sharing.
For this, we use repositories to share the source code of jobs together with corresponding runtime data on previous executions and their context.
Further, we developed performance models that account for the individual execution contexts of different, globally distributed users.

\begin{figure}[htb!]
    \includegraphics[height=0.85\textheight, width=\columnwidth, keepaspectratio]{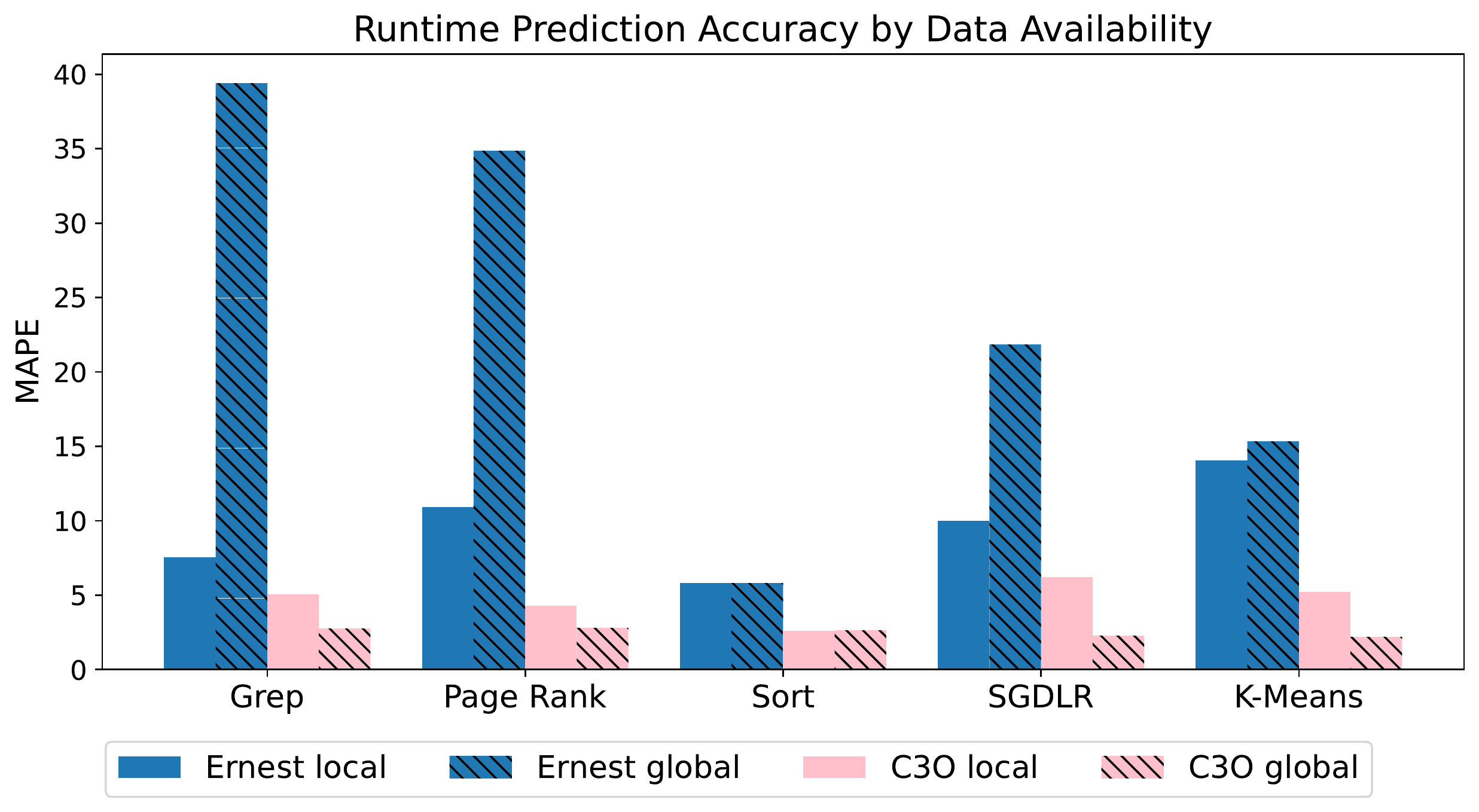}
    \caption{Using runtime metrics of local and global historical job executions to predict the runtime of a newly submitted processing job in a previously unseen execution context. The model performance is measured by mean absolute percentage error (MAPE). Results of~\cite{will2021c3o}.}
    \label{fig:results_c3o}
\end{figure}

The performance of the C3O runtime predictor\footnote{C3O runtime predictor and detailed evaluation:\\\href{https://github.com/dos-group/c3o}{https://github.com/dos-group/c3o}} is presented in~\autoref{fig:results_c3o}, where it is compared to Ernest~\cite{venkataraman2016ernest}.
Both predictors were evaluated on a dataset of 930 unique Spark 2.4 jobs that were executed on Amazon EMR, consisting of different numbers of AWS machines of categories \emph{c}, \emph{m}, and {r}, and sizes \emph{large}, \emph{xlarge}, and \emph{2xlarge}, which represent different allocations of memory and vCPUs per VM.
The jobs in this dataset further cover a variety of algorithm parameters (e.g.\ \emph{k} in K-Means) and key dataset characteristics (e.g.\ the number of features and observations in Linear Regression).
The full experimental setup is documented in~\cite{will2020towards}.

The C3O predictor already outperforms Ernest when trained only on training data stemming from the same, \emph{local}, context.
This difference is exacerbated when the predictors are trained with \emph{global} runtime data, since C3O is context-aware and can make good use of the information, while Ernest cannot.
Hence, we observe that our collaborative approach with context-aware runtime prediction models outperforms traditional single-user approaches, especially when shared runtime metrics are available.

In another work extending the C3O system, we presented a way to minimize storage and transfer costs by reducing the training data while retaining model accuracy~\cite{will2021training}.

\paragraph{Limitations and Future Work}

A major limitation of the approach we presented in C3O is that it only works for well-established jobs (like Grep or K-Means).
We will therefore work on performance models and cluster configuration methods that do not rely on the particular job having been previously executed elsewhere.
To solidify the sharing angle of our methods, we will work on approaches for data validation and establishing trust among users of collaborative systems that rely on shared runtime data.

\subsection{Performance Prediction with Context-Aware and Reusable Models}
\label{sec:results_performance_prediction}

In the following, we present first results on performance models that are able to detect and leverage differences in the execution context of jobs and are thus more reusable.

\paragraph{Results Overview}

\begin{figure}[htb!]
    \centering
    \includegraphics[width=\columnwidth]{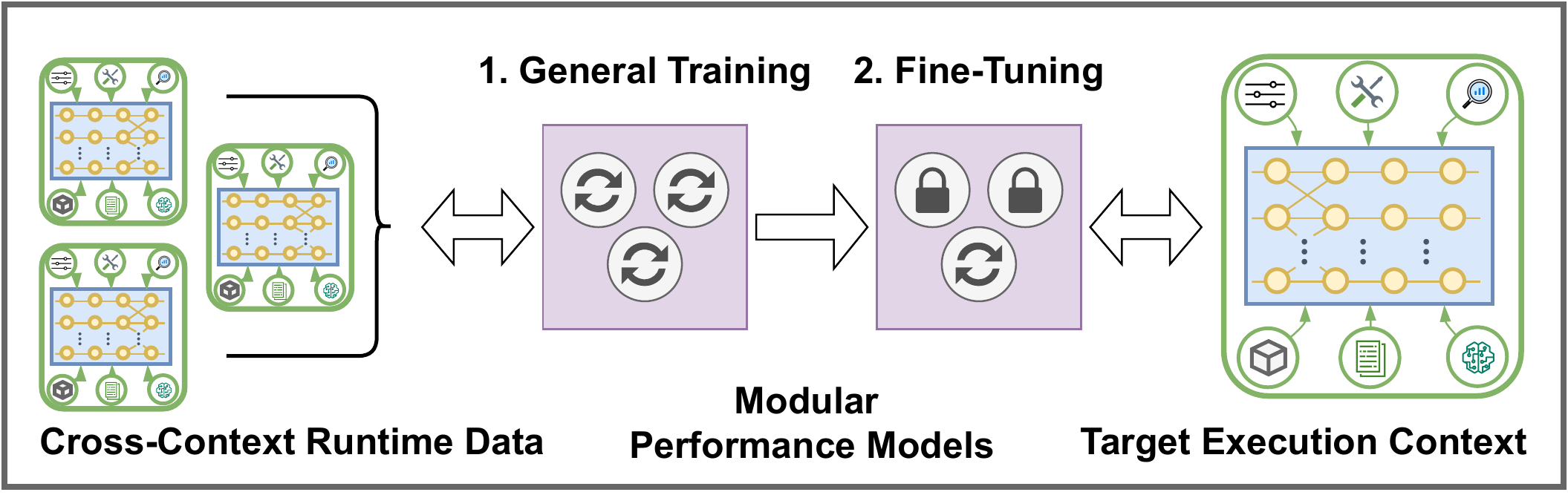}
    \caption{Context-awareness and reusability of performance models for utilization of runtime data across contexts.}
    \label{fig:results_performance_prediction}
\end{figure}

The previous subsection underlined the benefits of data and model sharing across users.
It also demonstrated that a proper representation of a job execution in form of data measurements has advantages.
To this end, we achieved initial results on performance prediction with context-aware and reusable models~\cite{verbitskiy2018cobell, scheinert2021bellamy}.
More specifically, we developed first models which are able to differentiate and leverage different execution contexts, going beyond fairly general performance models and automatic model selection as seen in our earliest work on the topic, Bell~\cite{thamsen2016bell}.
This idea is depicted in~\autoref{fig:results_performance_prediction}.

\begin{figure}
    \centering
    \includegraphics[width=\columnwidth]{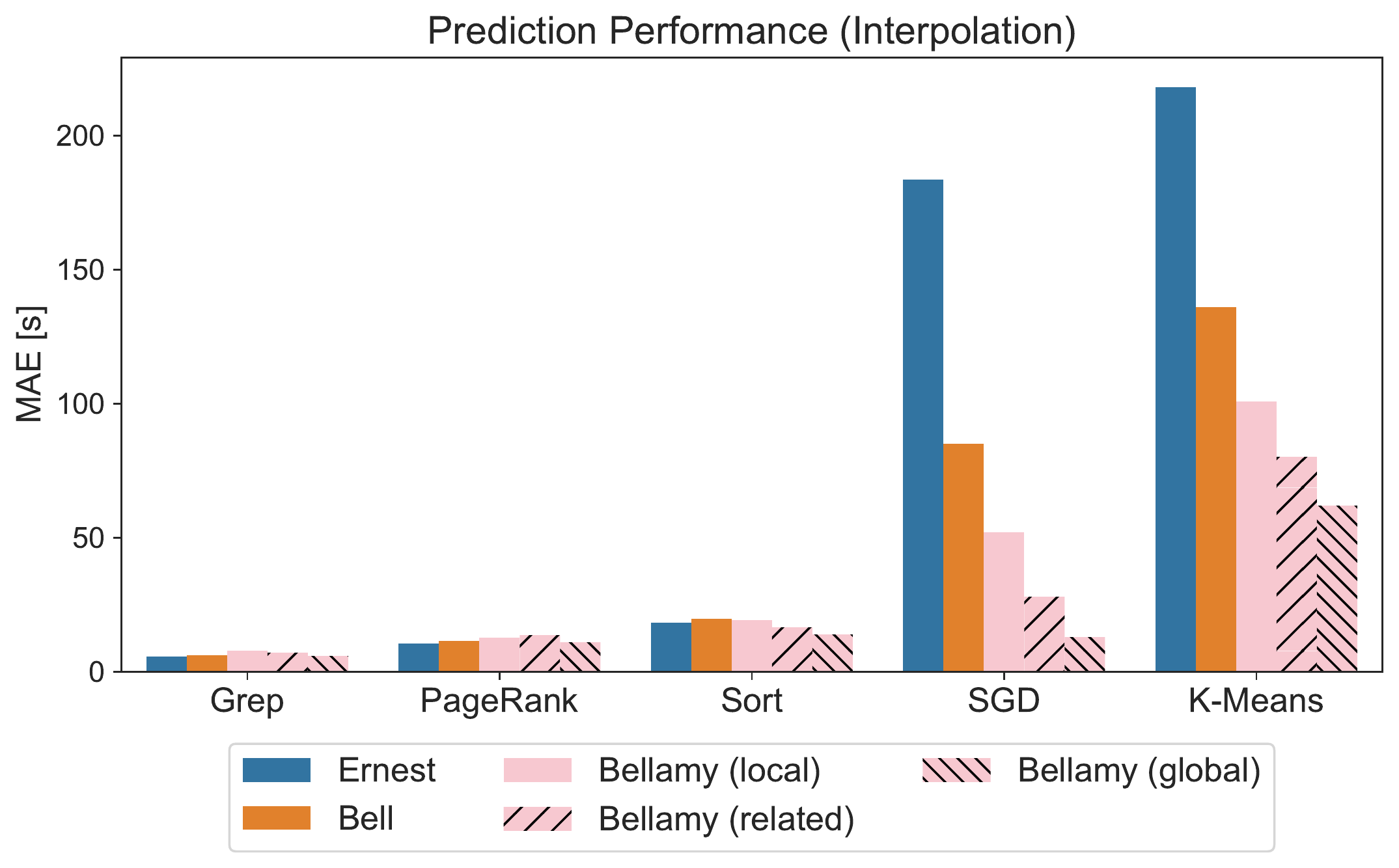}
    \caption{Reusable models can leverage related historical execution data for improved runtime prediction accuracy. The model performance is measured by mean absolute error (MAE) across various configurations. Results of~\cite{scheinert2021bellamy}.}
    \label{fig:results_performance_prediction_bellamy}
\end{figure}

\paragraph{Central Findings}

Our first efforts~\cite{thamsen2016bell,verbitskiy2018cobell} encompassed the utilization of various regression models with comparably few parameters.
While fast to train, these models are too simple to sufficiently capture the execution context of dataflow jobs, leading to large estimation errors.
More recently, we therefore investigated the applicability of neural networks for performance estimation and found that they can improve previous results by a significant margin~\cite{scheinert2021bellamy}.
Specifically, we trained a multi-component neural network on data originating from various execution contexts.
At its core, our neural network architecture Bellamy\footnote{Model architecture and detailed instructions:\\\href{https://github.com/dos-group/bellamy-runtime-prediction}{https://github.com/dos-group/bellamy-runtime-prediction}} utilizes an auto-encoder for encoding and exploiting descriptive properties of the enclosing execution context.
This approach effectively enables the reuse of data from various contexts and a better approximation of a job's scale-out behavior, hence leading to improved prediction results.
For any encoded resource configuration as input, Bellamy predicts a runtime value, which can in turn be used to select the best candidate configuration according to user-defined objectives and runtime constraints.

The full experimental setup can be found in~\cite{scheinert2021bellamy}, though in short, we utilized the C3O dataset and investigated among other things the interpolation and extrapolation capabilities of our approach under varying data availability for model pre-training using random sub-sampling cross-validation and various concrete model configurations.
\autoref{fig:results_performance_prediction_bellamy} shows the interpolation results for various dataflow jobs, where the advantage of our approach over comparative methods is especially evident for jobs with presumably non-linear scale-out behavior.
It can be seen that even hardly related data can improve the prediction performance for the execution context at hand.
We also find that this generally mitigates the cold-start problem through the incorporation of knowledge from historical workload executions.
Moreover, such neural network models can be pre-trained and fine-tuned, and are thus reusable across contexts.

\paragraph{Limitations and Future Work}

Currently, our approach mainly leverages textual properties, which are often difficult to interpret and compare.
Thus, a promising direction is the concretization of textual properties of a job execution, i.e. by measuring and including appropriate metrics.
Moreover, there is room for improvement with regards to efficient model re-trainings from scratch, as well as appropriate stopping criterions in case of limited available training data.
We also plan to incorporate further aspects of the execution context (such as explicit information about the underlying infrastructure) into prediction models systematically in the future.

\subsection{Runtime Adjustments of Performance Models and Resource Configurations}
\label{sec:results_runtime_adjustments}

Next, we present concrete techniques to realize dynamic changes to cluster configurations for data-parallel distributed processing jobs.

\paragraph{Results Overview}

\begin{figure}
    \centering
    \includegraphics[width=\columnwidth]{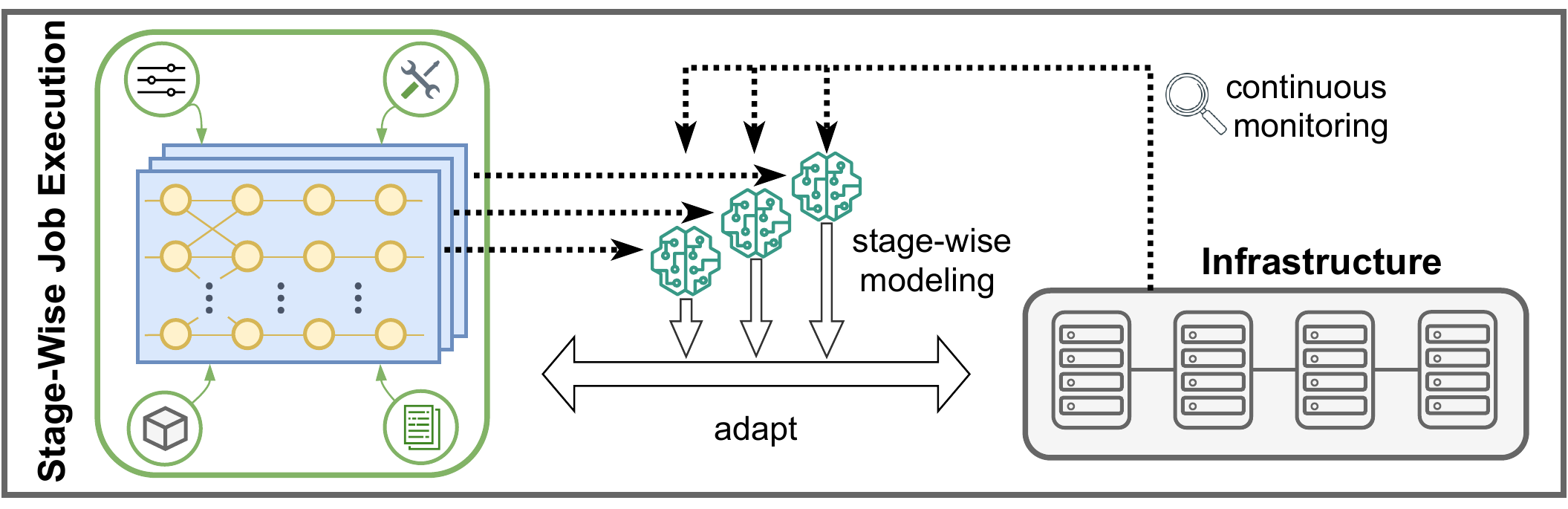}
    \caption{Continuous adjustments of performance models and resource configurations along the structure of jobs.}
    \label{fig:results_runtime_adjustments}
\end{figure}

The aforementioned results optimize initial resource allocations.
Going beyond this, we worked on mitigating many dynamic effects that cannot be foreseen accurately, such as changing input data distributions or interference with co-located jobs.
To continuously facilitate efficiency, while meeting provided runtime targets, we have developed approaches for dynamic adjustments of performance models and resource configurations~\cite{thamsen2017ellis,scheinert2021enel}.
The general idea is depicted in~\autoref{fig:results_runtime_adjustments}.

\begin{figure}
    \centering
    \includegraphics[width=\columnwidth]{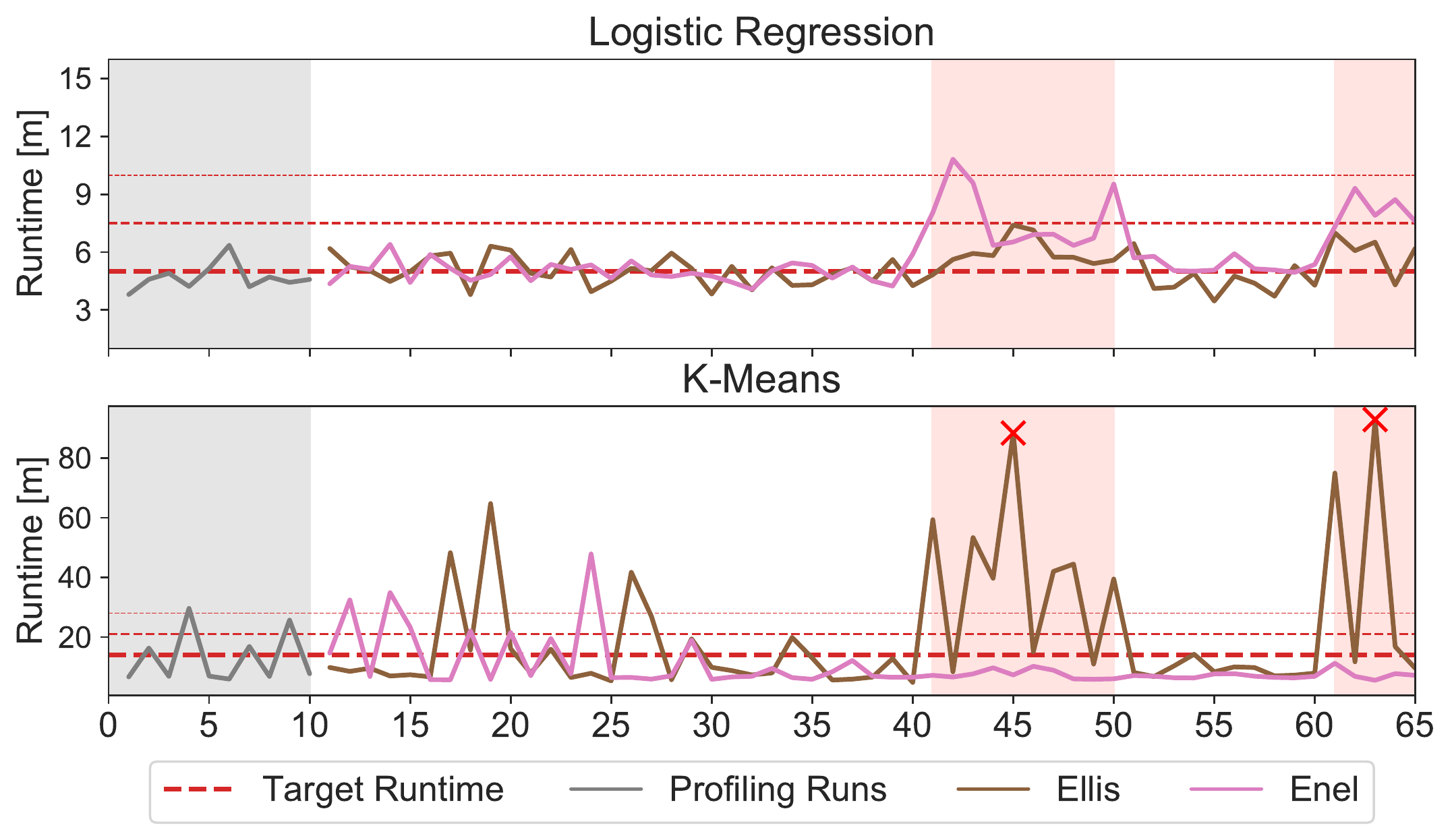}
    \caption{Dynamic job scaling in light of normal (white background) and unexpected behavior (highlighted in red). Deviations of the respective runtime targets are less critical for our global graph model. Results of~\cite{scheinert2021enel}.}
    \label{fig:results_runtime_adjustments_enel}
\end{figure}

\paragraph{Central Findings}

Worth discussing here is that we implemented dynamic adjustments along the structures of iterative jobs, namely synchronization barriers between subsequent job iterations to re-assess and, if needed, change configurations.
In particular, we build upon our idea initially proposed with Ellis~\cite{thamsen2017ellis}, where we introduced a dynamic horizontal scaling method for distributed dataflow jobs with runtime targets and leveraged the fact that iterative jobs can be logically dissected into many individual stages.
Yet, in contrast to training an ensemble of stage-specific, specialized performance prediction models, we turned to employ a single global graph model with Enel~\cite{scheinert2021enel}, which is trained on the entire available execution data.
Annotated with descriptive properties and collected monitoring metrics, the directed acyclic graph of tasks in individual stages, as well as the graph of all stages on a meta-level, is leveraged and exploited by our global model for proper rescaling recommendations via runtime prediction and subsequent prediction-based configuration ranking.
Moreover, in contrast to Ellis, our newer method Enel does not simply employ heuristics to assess the trade-off between rescaling and its overhead, but learns the expected overhead in a data-driven manner and incorporates this into the decision-making process through consideration of the aforementioned properties and graph structures.

The full experimental setup can be found in~\cite{scheinert2021enel}, where we compare Enel\footnote{Implementation and experiment details:\\\href{https://github.com/dos-group/enel-experiments}{https://github.com/dos-group/enel-experiments}} against Ellis on a selection of four commonly employed iterative Spark jobs running in a commodity cluster of 50 machines.
Each job is executed 65 times, where for some executions we simulated anomalous behavior by randomly injecting failures into Spark executors.
Moreover, we investigated the scale-out range from 4 to 36 Spark executors.
We found that this single model is more robust and reusable across the stages of dataflow jobs.
Though requiring a sufficient amount of data which manifests in a longer profiling phase, our global graph models tend to better capture the enclosing execution context in the long run.
It is superior in detecting and mitigating anomalous execution behavior as shown in~\autoref{fig:results_runtime_adjustments_enel}, and requires only a single generalized model instead of a multitude of specialized ones.

\paragraph{Limitations and Future Work}
A major limitation of our current approach is the assumption of jobs being executed in isolation, ignoring the potential interference with co-located jobs.
Moreover, the selection and appropriate representation of monitoring metrics remains a challenge.
In the future, we further intend to integrate forecasting methods into our approaches to enable pro-active dynamic resource configurations and we will work on more accurately identifying points in time that are especially suitable for performance model updates.
In addition, we want to investigate how our approach to cluster resource configuration can be used in combination with similar ideas for indexing~\cite{ding2019ai} or query processing~\cite{leis2021towards}.

\section{Summary}
\label{sec:summary}

In this paper, we charted our work towards a more collaborative approach to cluster configuration for distributed data-parallel processing.
We envision for our approach that users share runtime data and performance models for their processing jobs across different execution contexts.
Central to enabling this collaborative vision will be the following building blocks: (1) methods for taking advantage of the similarity of computational resources and jobs for performance predictions, (2) techniques to reuse performance model components across different contexts, as well as (3) strategies to adjust prediction models and cluster configurations dynamically.
Though we have presented major results for each of these three building blocks, more work is needed to fully realize our vision of a more collaborative cluster configuration.
We therefore plan to continue this line of work, aiming to help more users take advantage of distributed data-parallel processing systems and thereby democratize access to large commodity compute clusters.

\backmatter





\bmhead{Funding}

This work has been supported through grants by the German Ministry for Education and Research (BMBF) as BIFOLD (grant 01IS18025A) and the German Research Foundation (DFG) as FONDA (DFG Collaborative Research Center 1404).

\bibliography{sn-bibliography}

\end{document}